\documentclass[12pt]{article}

\usepackage{sbc-template}
\usepackage{graphicx,url}
\usepackage[utf8]{inputenc}

\title{Teaching Practically Relevant Research Problem Formulation in Software Engineering with Lean Research Inception}

\author{Anrafel Fernandes Pereira\inst{1}\inst{2}, Tatiane Ornelas\inst{1}, Allysson Allex Araújo\inst{3}, and \\ Marcos Kalinowski\inst{1}}

\address{Pontifical Catholic University of Rio de Janeiro - PUC-Rio, Rio de Janeiro, Brazil.
\nextinstitute
  University of Vassouras, Vassouras, Rio de Janeiro, Brazil.
\nextinstitute
    Federal University of Cariri (UFCA), Juazeiro do Norte, Ceará, Brazil
  \email{\{afpereira,talves,kalinowski\}@inf.puc-rio.br,}  \email{allysson.araujo@ufca.edu.br} 
}

\begin{document} 

\maketitle
\begin{abstract}
[Background] Well-formulated Software Engineering (SE) research problems are essential for bridging the gap between industry-academia. Lean Research Inception (LRI) aims to support this activity. [Goal] Apply LRI to support SE students in formulating practice-aligned research problems. [Method] We conducted a case study with 60 students and 7 faculty advisors of a Brazilian university. [Results] Students reported benefits in reasoning (60\%), clarity and definition (61.7\%), contextualization (60\%), and communication (50\%). Advisors also observed clearer and more structured problems (57.1\%) with a high recommendation rate (85.7\%). [Conclusion] LRI can be a promising approach to support practice-aligned research problem formulation in SE education.
\end{abstract}

\section{Introduction}
Preparing future software engineers to deal with real-world challenges requires educational approaches that integrate academic training with industry needs~\cite{romao2024agile} ~\cite{molleri2024teaching}. The literature in Software Engineering (SE) Education has emphasized the need to bridge this gap. Several studies indicate that this mismatch is partly due to the difficulty of exposing undergraduate students to realistic problems and supporting them in effectively articulating theory, context, and practical application~\cite{greco2025learning}. In this scenario, final year projects and capstone projects have been widely adopted as strategies to integrate technical knowledge and develop essential professional skills~\cite{da2024undergraduate, codabux2024software, goelzer2025s}, such as teamwork, communication, and decision-making.

Learning to formulate well-defined research problems is an important step in the education of software engineers, especially when the goal is to bring academic research closer to industry needs. The low practical relevance of many studies in SE has been attributed to research problems that overlook the real needs of the industry or rely on overly simplified assumptions about the professional context~\cite{garousi2020}. Despite the recurring concern with practical relevance, the literature has prioritized the evaluation of research outcomes~\cite{molleri2023,petersen2024}, while giving less attention to the initial stage of problem formulation, precisely where relevance begins to take shape. Supporting students at this early stage can be a differentiator in their academic training.

In this context, Lean Research Inception (LRI)~\cite{PaperEASE2025} emerges as a structured approach to support the formulation and early assessment of practically relevant research problems in SE. This study explores the application of LRI as a pedagogical strategy to teach research problem formulation in SE, emphasizing the formulation of practically relevant problems from the outset. To this end, we applied the Problem Vision board~\cite{PaperWSESE2026}, one of the core instruments of LRI, in a Capstone Project course within the undergraduate SE program across three campuses of the University of Vassouras\footnote{\url{https://univassouras.edu.br}}. Students were trained, guided, and used LRI to formulate their research problems with the support of their faculty advisors. Perceptions were collected through questionnaires voluntarily completed by 60 students and 7 advisors involved in the study.

The results show that LRI contributed to students’ learning. Most students reported improvements in organizing their reasoning (61.7\%), understanding and applying the board’s attributes (60\%), and clearly communicating their research problems (50\%). LRI was also perceived as easy to use (60\%) and valuable for future projects (58.3\%). Faculty advisors noted greater clarity in the structure of the formulated problems (57.1\%), support for aligning them with practical needs (42.9\%), and improved communication during supervision (57.1\%). Ease of use was highlighted by 71.4\% of the advisors, and 85.7\% expressed strong intention to recommend the approach. Regarding the contributions, this study positions research problem formulation as a teachable and learnable competence in SE education. It introduces and empirically evaluates Lean Research Inception (LRI) as a pedagogical strategy to support this early and often overlooked phase.

Unlike previous studies that focus on project execution or the use of empirical methods, this study emphasizes the initial formulation of practice-aligned research problems as an educational objective. The results show that LRI can help avoid vague or disconnected problem definitions and foster early alignment with real-world demands, reinforcing its potential as a scalable and reusable approach in SE education. This study can contribute to SE education in three main ways. First, it frames research problem formulation as a teachable competence in SE education. Second, it explores how LRI can be used pedagogically to scaffold students during the early stages of research design in Capstone courses. Third, it provides empirical evidence from a real educational context involving both students and faculty advisors, highlighting perceived learning benefits.

\section{Related Work}
Practice-based projects have been widely adopted to bring students closer to real-world scenarios. Courses structured as academic software factories report advances in professional maturity and in the development of technical, socio-emotional, and ethical competencies by engaging students in real projects with external stakeholders ~\cite{da2024undergraduate}.
%Similarly, programs based on university–industry partnerships promote the progressive insertion of students into real production environments, contributing to the development of technical and behavioral skills and to reducing the gap in this scenario~\cite{goelzer2025s}.

Other studies explore educational models that combine industry–academia collaboration with active learning methodologies, such as the integration of Lean R\&D ~\cite{kalinowski2020lean} with Problem-Based Learning (PBL)~\cite{romao2024agile}. The results show high acceptance of the approaches and improved technical competencies and business understanding, driven by engagement with real-world problems. Additionally, research has investigated the teaching of research design and empirical methods as fundamental competencies in SE education. Reports of SE Research Methods courses and experiences based on Active Learning indicate high levels of student satisfaction and gains in knowledge, confidence, and motivation, although challenges related to clarity of objectives and the development of critical thinking are still observed \cite{vilela2024empowering, greco2025learning}. Molléri and Petersen (2024) stands out for proposing a structured approach to support students in conducting research in SE, using guides, frameworks, and templates~\cite{molleri2024teaching}. Initiatives like this aim to facilitate the identification of relevant problems, the formulation of research questions, and the selection of appropriate methods, promoting a bridge between rigor and relevance.

Despite these advances, most studies focus on project execution or the teaching of empirical methods, devoting less attention to the initial phase of research problem formulation, especially in the context of Capstone Project courses. It is within this gap that the contribution of this work is positioned, by investigating the application of Lean Research Inception as a pedagogical strategy to support students in the formulation of practice-aligned research problems.

\section{Lean Research Inception (LRI)}
Lean Research Inception~\cite{PaperEASE2025} aims to support Software Engineering (SE) researchers in the formulation and initial assessment of research problems aligned with professional practice. The approach seeks to address one of the main challenges in SE research: the misalignment between academically investigated problems and the real needs of industry. LRI is grounded in agile principles and methodologies such as Design Thinking~\cite{plattner2009}, Lean Startup~\cite{ries2011}, and Lean Inception~\cite{caroli2017}, and emphasizes early collaboration between researchers and industry practitioners. Its core objective is to integrate different perspectives from the early stages of the research process, fostering the construction of research problems that are clearer, better contextualized, and more relevant to real-world scenarios.

\begin{figure}[h!]
    \centering
    \includegraphics[width=0.75\linewidth]{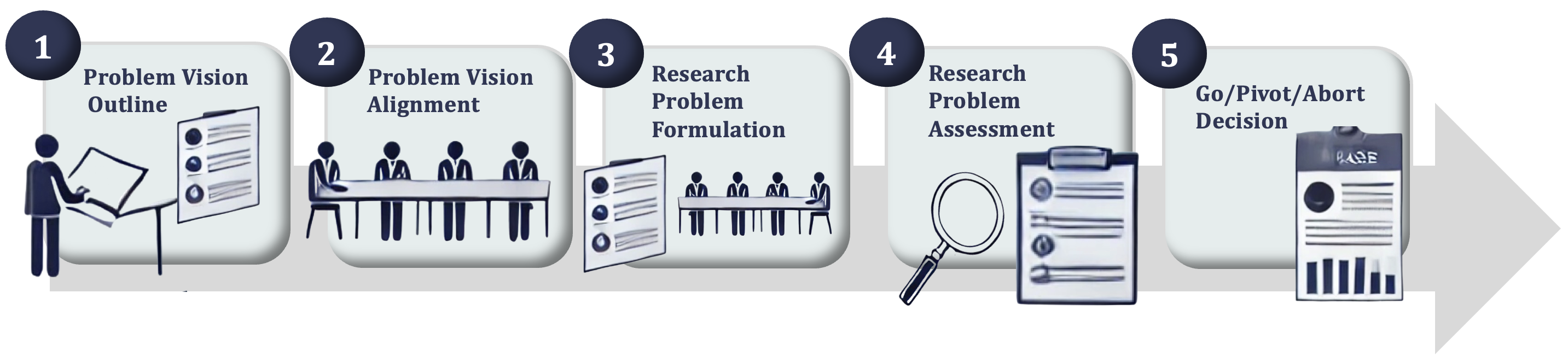}
    \caption{Lean Research Inception Overview}
    \label{fig:lri_overview}
\end{figure}

The approach is structured into five sequential phases: \textit{1 - Problem Vision Outline}, \textit{2 - Problem Vision Alignment}, \textit{3 - Research Problem Formulation}, \textit{4 - Research Problem Assessment}, and \textit{5 - Go/Pivot/Abort Decision}. To support its application, LRI provides a public visual template\footnote{https://miro.com/miroverse/lean-research-inception-template/}, available to the SE community, which enables collaborative and traceable research problem formulation. Figure~\ref{fig:lri_overview} presents an overview of the approach and its phases. The first three phases of the LRI are dedicated to formulating the research problem and constitute the focus of this study.

\begin{figure}[h!]
    \centering
    \includegraphics[width=0.7\linewidth]{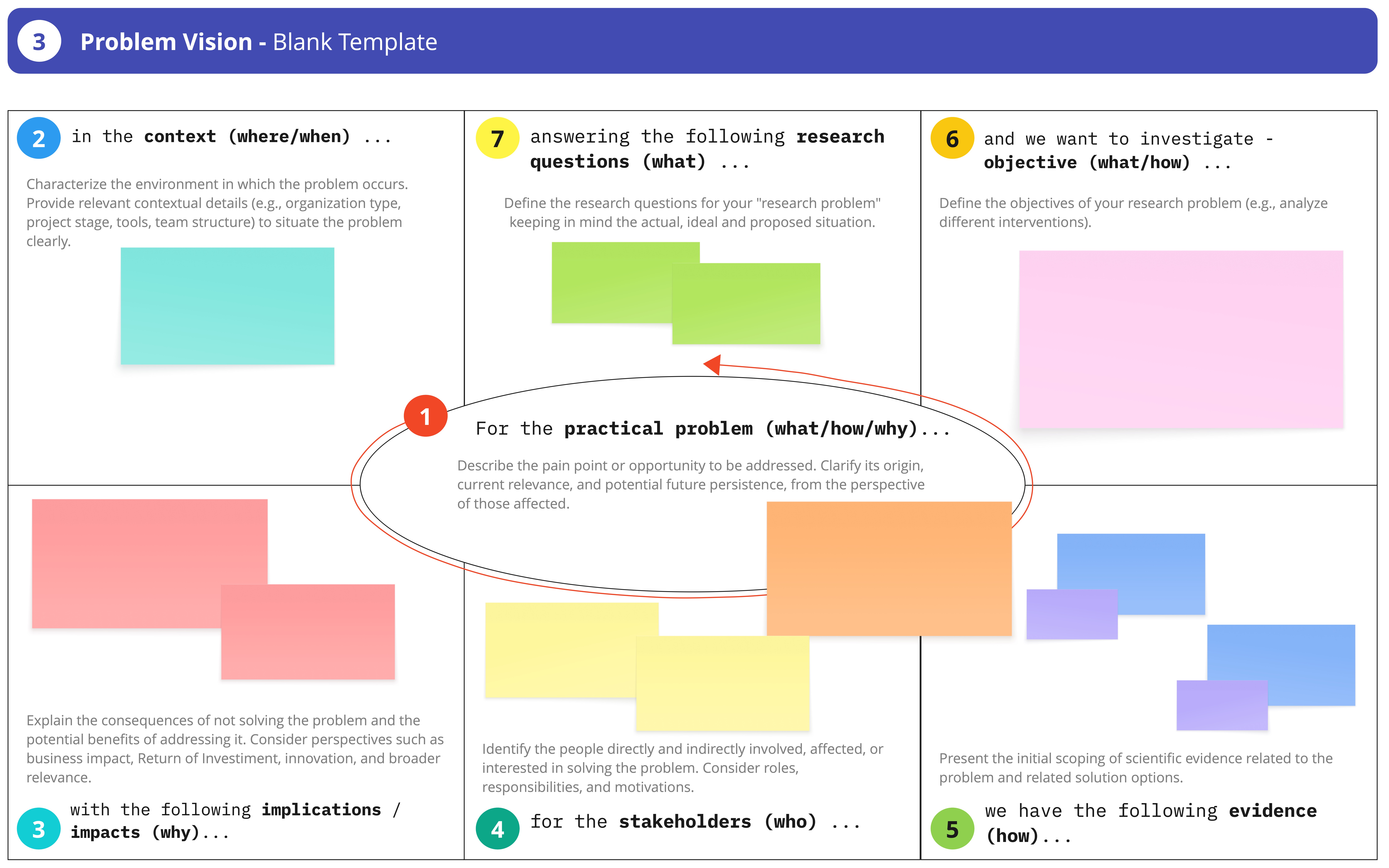}
    \caption{Problem Vision board}
    \label{fig:problem_vision}
\end{figure}

\textbf{Phase 1 – Problem Vision Outline:} In this phase, researchers create an initial version of the research problem using the Problem Vision board. The problem is structured based on seven attributes: \textit{practical problem}, \textit{context}, \textit{implications/impacts}, \textit{stakeholders}, \textit{evidence}, \textit{objective}, and \textit{research questions}. Figure~\ref{fig:problem_vision} illustrates the Problem Vision Board and the organization of its attributes~\cite{PaperWSESE2026}.

\textbf{Phase 2 – Problem Vision Alignment:} In this phase, industry practitioners are invited to collaborate with researchers in a problem alignment process. The previously elaborated Problem Vision is discussed in a structured manner, allowing assumptions to be challenged, practical experience to be incorporated, and the relevance of the problem to be validated. Iterative adjustments are made directly on the board, focusing on clarifying ambiguities, strengthening evidence, and aligning objectives and research questions.

\textbf{Phase 3 – Research Problem Formulation:} Based on the alignment achieved, the research problem is consolidated and documented. The attributes of the Problem Vision are revisited to ensure clarity, consistency, and alignment. The outcome of this phase is a well-defined, practice-oriented research problem shared among the involved stakeholders, serving as a reference for subsequent phases of the LRI.

\textbf{Phase 4 (Research Problem Assessment)} and \textbf{Phase 5 (Go/Pivot/Abort)} focus on evaluating and making strategic decisions regarding the formulated problem, requiring broader stakeholder involvement.

\section{Study Design}
In this study, students were instructed to assume the role of Software Engineering (SE) researchers, using the Problem Vision board to structure their research problems during \textbf{Phase 1 - Problem Vision Outline}. Subsequently, these problems were discussed and refined with their faculty advisors, who acted as representatives of professional practice during the \textbf{Phase 2 - Problem Vision Alignment} and \textbf{Phase 3 - Research Problem Formulation} phases. This dynamic allowed the advisors’ practical experience to be incorporated into the formulation process, promoting greater clarity, contextualization, and alignment of the research problems developed by the students. Considering the educational context of Capstone Project courses and the main objective of this study, including Phases 4 and 5 would have shifted the focus of the study and introduced additional variables not aligned with its goals. Restricting the study to the research problem formulation phases allowed a more focused analysis of the pedagogical contributions of LRI.

This study was conducted following methodological guidelines for case studies in SE~\cite{runeson2012case}. Such an approach is particularly suitable for investigating the application of methods, processes, or approaches in real-world contexts, where the phenomenon under study cannot be dissociated from the environment in which it occurs. The choice of a case study is justified by the objective of understanding how the LRI behaves when applied in an educational context, more specifically in SE education, allowing the observation of perceptions, impacts, and emerging challenges based on the experiences of students and faculty advisors. The study adopts an interpretive perspective, combining quantitative and qualitative data to provide a comprehensive view of the contribution of the approach in supporting the formulation of practice-aligned research problems.

\subsection{Goal and Research Questions}
We defined the goal of this study using the goal definition template of the Goal Question Metric (GQM) paradigm~\cite{basili1988} as follows: ``\textbf{Analyze} the application of Lean Research Inception
\textbf{for the purpose of} characterization
\textbf{with respect to} the perceived support for practice-aligned research problem formulation, considering support for reflection on the problem, clarity, structure, alignment with practice, communication, ease of use, and adoption or recommendation of the approach 
\textbf{from the point of view of} Software Engineering (SE) students and faculty advisors \textbf{in the context of} Capstone Project courses across three campuses of a Brazilian university''. To investigate this goal, we formulate the following research questions (RQs):
\begin{itemize}
    \item \textbf{RQ1:} How do SE students perceive the support provided by Lean Research Inception for practice-aligned research problem formulation, considering aspects such as reflection on the problem, clarity and definition, ease of communication, ease of use, and intention to adopt the approach in other projects?
    \item \textbf{RQ2:} How do faculty advisors perceive the support provided by Lean Research Inception for practice-aligned research problem formulation, considering aspects such as clarity and structure, alignment with practical needs, ease of communication, ease of use, and intention to recommend the approach?
\end{itemize}

\subsection{Case and Subject Selection}
The case study was conducted in the context of the Capstone Project course in the Software Engineering (SE) program at the University of Vassouras, Rio de Janeiro (RJ) - Brazil. The course is offered in a distance learning format, with a workload of 60 hours, and is carried out in the final stage of the Bachelor's Degree in SE. Its main objective is to support students in defining a research problem and developing a SE solution directly associated with that problem. In this context, students are guided to articulate theoretical foundations, application context, and technical decisions throughout the project. Projects may be developed individually or in groups of up to four students, enabling both individual and collaborative experiences and fostering the development of technical, analytical, and teamwork skills. In addition, each student or group may choose their project advisor, who accompanies and guides the development of the work throughout the entire course.

All students enrolled in the selected course cohorts were invited to participate in the study, following a convenience sampling strategy. Participation in the questionnaire was voluntary and anonymous, and students were informed that it would not affect their academic evaluation. The study involved three course cohorts distributed across three different campuses of the University of Vassouras: Vassouras/RJ, Maricá/RJ, and Saquarema/RJ. In total, 60 students regularly enrolled in the course and seven faculty advisors responsible for supervising the Capstone Projects voluntarily participated in the study. All participants had access to the same materials, general guidelines, and activities, ensuring consistency in the application of the approach. Participation in the study, particularly in completing the evaluation questionnaires, was voluntary and had no academic impact on those who chose not to participate.

\subsection{Instrumentation}
Lean Research Inception was integrated into the course in a structured manner through a set of pedagogical and scientific instruments made available along the Learning Path in the Learning Management System (LMS). The main instruments used included instructional materials in the LMS, such as: explanatory texts detailing the concepts and principles of LRI; introductory and guidance videos on the application of the approach; a complete practical example of applying the Problem Vision board; the Problem Vision board template, provided in a collaborative digital format through the Miro platform and as static images for reference and offline use; and a scientific article~\cite{PaperEASE2025} describing the LRI approach, used as conceptual support material. 

In addition, synchronous remote sessions were conducted via Microsoft Teams, during which students were introduced to the theoretical foundations of the approach, a practical example of its application, and guidelines for using the board in the context of the Capstone Project course. Students also benefited from asynchronous support provided by faculty advisors throughout the application period, allowing for continuous clarification of questions. All materials are available in our open science repository~\footnote{\url{https://zenodo.org/records/19074876}}~\cite{zenodoRepository}, including some examples of research problems formulated by students using the Problem Vision board.

\subsection{Data Collection and Analysis Procedures}
Data collection was conducted after the students completed the research problem formulation stage. Two distinct questionnaires were used: one directed at students and the other at faculty advisors. Prior to data collection, participants were informed about the study objectives and the voluntary nature of participation. Consent was obtained through the questionnaire, and no personal identification data were collected, ensuring anonymity.

The student questionnaire was structured around clarity and structuring of the research problem, contextualization, practical applicability, communication of the problem, and interaction with advisors. It included five-point Likert scale questions (e.g., \textit{``Did the use of the Problem Vision board help you reflect more effectively on the research problem addressed in your project?''}) and open-ended questions (such as ``\textit{What were the main benefits you perceived from using the Problem Vision board?}''). The faculty advisor questionnaire followed a similar structure, with questions focused on their perception of the quality of the formulated problems, clarity and practical feasibility of the proposals, and their intention to adopt the approach in future projects. Quantitative data were analyzed using descriptive statistics (frequencies and percentages). Open-ended responses were analyzed through qualitative content analysis using open coding to identify recurring patterns and complementary insights. To strengthen the interpretation of results, we triangulated quantitative questionnaire data with qualitative evidence from open-ended responses, providing a more comprehensive understanding of participants’ perceptions.

\section{Results}
\subsection{Case and Subject Description}
The study involved 60 students in their final year of undergraduate studies in SE and seven faculty advisor responsible for the Final Year Projects across the three participating campuses. Table~\ref{tab:advisors} presents the characteristics of the faculty advisors participating in the study. Unfortunately, we did not have participation from advisors from all campuses in this study, since participation was voluntary. All students had access to the same materials, guidelines, and instruments provided through the LMS, as well as a synchronous presentation of the approach and recorded sessions for those unable to attend live. The period allocated for formulating the research problem was approximately three weeks, during which students could interact asynchronously with their advisors. At the end of this period, students submitted their formulated research problems, and both students and advisors were subsequently invited to complete the questionnaire evaluating the approach.

\begin{table}[h!]
\centering
\caption{Characterization of the Faculty Advisors}
\resizebox{\textwidth}{!}{%
\begin{tabular}{|c|l|l|l|l|}
\hline
\textbf{ID} & \textbf{Campus} & \textbf{Academic Training} & \textbf{Years of experience} & \textbf{Guidelines for Final Course Projects} \\ \hline
P1 & Vassouras & Master's degree & More than 14 years & More than 21 guided projects \\ \hline
P2 & Vassouras & Specialist & More than 3 years & Fewer than 10 guided projects\\ \hline
P3 & Maricá & Doctorate & More than 2 years & Fewer than 10 guided projects\\ \hline
P4 & Vassouras & Master's degree & More than 6 years & Between 11 and 20 guided projects\\ \hline
P5 & Vassouras & Master's degree & More than 4 years & Fewer than 10 guided projects\\ \hline
P6 & Vassouras & Doctorate & More than 11 years & More than 21 guided projects\\ \hline
P7 & Vassouras & Specialist & More than 1 year & Fewer than 10 guided projects\\ \hline
\end{tabular}%
}
\label{tab:advisors}
\end{table}

The results are organized around two research questions. RQ1 analyzes SE students' perception of the support provided by the Lean Research Inception, and RQ2 analyzes the advisors' perception. The analysis combines quantitative data (see Figure~\ref{fig:chart_students} for RQ1 and Figure~\ref{fig:chart_advisors} for RQ2) with qualitative insights from open-ended responses, providing an integrated view of participants’ perceptions.

\subsection{Students' Perceived the support provided by Lean Research Inception (RQ1)}
\begin{figure}[h!]
    \centering
    \includegraphics[width=0.8\linewidth]{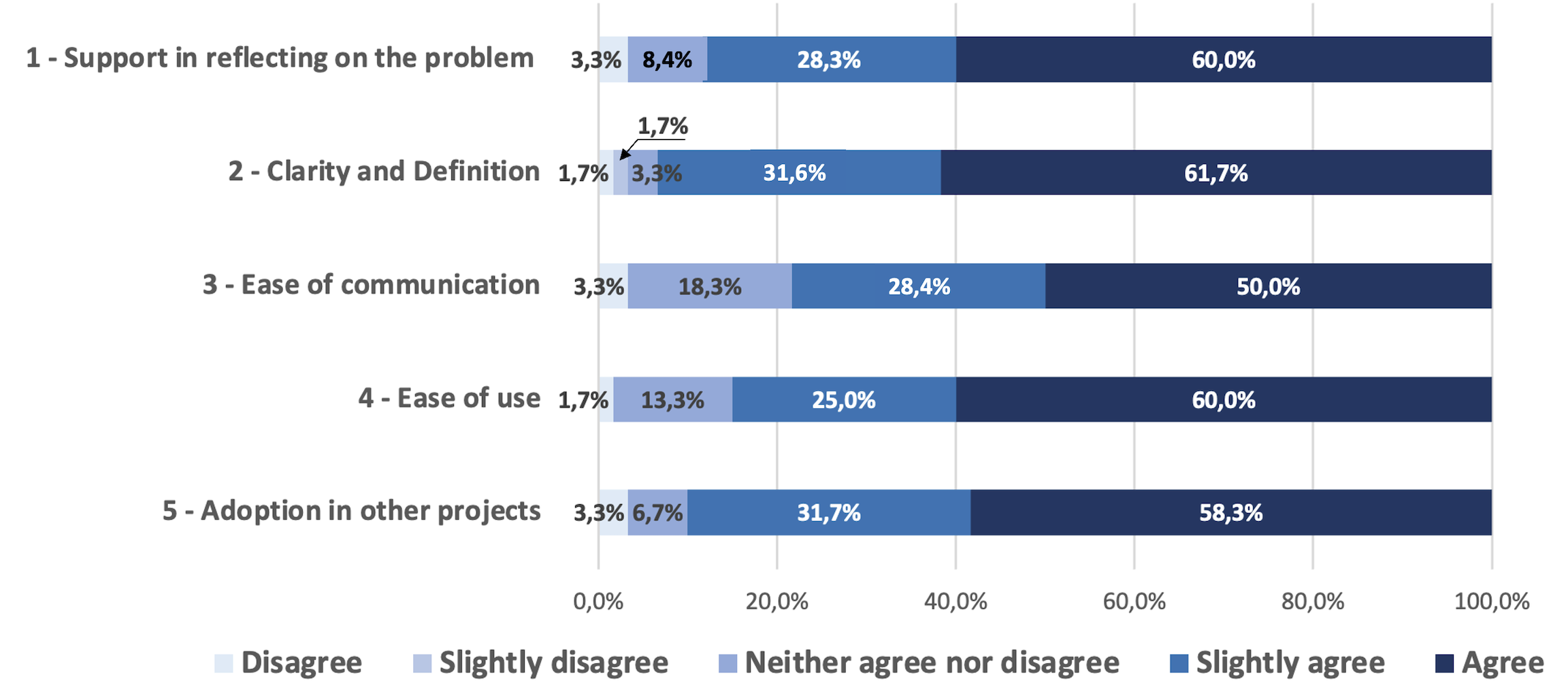}
    \caption{Results of the Students' Questionnaire (\%)}
    \label{fig:chart_students}
\end{figure}

The results of RQ1 focus on student perceptions. The analysis addresses support for reflection on the problem, clarity and definition, ease of communication, ease of use, and the intention to adopt the approach in other projects. Figure ~\ref{fig:chart_students} shows the results.

\textbf{Support for reflection on the research problem:} Most students reported positive perceptions of this topic. A total of 60\% of the students agreed and 28.3\% partially agreed that the use of the Problem Vision Board helped them reflect more effectively on their research problem, while only a small number expressed neutral or negative perceptions. These results indicate that the board played a relevant role in fostering a more reflective and structured thinking process during problem formulation. 

The qualitative feedback further reinforces this finding. Different students highlighted that the board helped them organize ideas and better understand the problem scope. For instance, P10 commented: \textit{``it helps a lot with clarity and organizing ideas''}, emphasizing the board’s contribution to structuring initial thoughts. Similarly, P4 noted that the board allowed them to \textit{``see the problem more clearly as a whole''}, suggesting support for holistic reflection. P1 also mentioned that the board \textit{``already helps a lot in organizing the problem''}, reinforcing its role as a cognitive aid during early problem formulation. In addition to positive feedback, some students provided suggestions that point to opportunities for improvement. P12 suggested that the visual organization of the board could be improved to make relationships between elements clearer, while P14 recommended the inclusion of a more explicit step-by-step guide to further support reflection, especially for students with less experience in research activities. These comments suggest that, although the board effectively supports reflection, additional instructional scaffolding could further enhance its pedagogical value.

\textbf{Clarity and definition of the research problem:} A total of 61.7\% of the students agreed and 31.6\% partially agreed that the Problem Vision board contributed to making their research problem clearer and better defined. Some students emphasized that the board helped transform vague defined ideas into more concrete problem statements. P1 emphasized that the board helped transform ideas into more concrete problem. P4 commented that the board helped them \textit{``understand better what the real problem was''}, suggesting that explicit attributes encouraged deeper analysis and refinement. Lastly, P10 also reinforced this point, stating that the board contributed to \textit{``greater clarity about what should actually be investigated''}. Some participants also identified opportunities for improvement related to clarity. P12 suggested that the visual layout could better emphasize the relationships between attributes, which could further support problem definition, especially in more complex projects. Additionally, P14 mentioned that clearer examples for each attribute could help students reach an adequate level of definition more quickly. These remarks indicate that, while the board effectively enhances clarity and definition, complementary guidance and visual refinements may further strengthen its impact.

\textbf{Ease of Communication:} 50\% of the students agreed and 28.4\% partially agreed that the Problem Vision board facilitated the communication of the research problem with peers and faculty advisors. However, compared to other dimensions, a higher number of neutral responses was observed, suggesting that the perceived communication benefits may vary depending on the interaction context or the way the board was used during discussions. Qualitative feedback helps explain these results. Some students emphasized that the visual and structured nature of the board supported clearer explanations of their ideas. For instance, P5 commented that \textit{``having everything visually organized makes it easier to explain the problem''}, highlighting the board’s role as a communication aid during supervision meetings. Similarly, P8 noted that the board helped to \textit{``organize thoughts before talking to the advisor''}, indicating that it also supported preparation for communication, not only the interaction itself. Other participants pointed out limitations or opportunities for improvement. P12 suggested that the communication benefits depend on familiarity with the board, noting that without proper explanation some elements may require additional clarification. Moreover, P14 mentioned that providing examples of how to present the board during discussions could further enhance its effectiveness as a communication tool. These comments suggest that, while the Problem Vision generally facilitates communication, its impact can be strengthened through clearer guidance on its use in collaborative interactions.

\textbf{Ease of use:} Most participants agreed (60\%) or partially agreed (25\%) that the board is easy to use and understand, indicating that the approach was accessible even for students at an early stage of research problem formulation and did not impose excessive cognitive effort. Several students explicitly stated that the board was intuitive and adequately structured for the purpose of the discipline. For example, P9 mentioned: \textit{``for this moment, it works well''} suggesting that the level of complexity was appropriate for the educational context. Similarly, P7 commented that the board was \textit{``simple and easy to understand''} highlighting its usability as a learning artifact. Some students also provided suggestions aimed at further improving usability. P14 suggested the inclusion of a more detailed step-by-step guide to support first-time users, while P12 mentioned that brief explanatory examples for each attribute could make the initial use even more straightforward. These comments indicate that, although the Problem Vision board is generally easy to use, additional instructional scaffolding could further enhance the user experience, particularly for students with less prior exposure to research activities.

\textbf{Intention to adopt the approach in other projects:} A total of 58.3\% of the students agreed and 31.7\% partially agreed that they would use the board again in future projects, indicating a high level of perceived usefulness and acceptance of the approach. Some students highlighted that the board added value by promoting organization, clarity, and structured thinking. For instance, P6 stated that the board \textit{``met expectations''} while P10 emphasized its contribution to clarity, noting that it helped make the problem more understandable and well organized. Likewise, P8 commented that the board was \textit{``useful and effective for structuring the problem,''} suggesting that its benefits were not limited to the specific discipline in which it was applied. Participants also discussed the circumstances that would favor the adoption of the approach in future contexts. P12 mentioned that the board would be particularly useful in projects with greater complexity or involving multiple stakeholders, while P14 suggested that continued use would be even more beneficial if accompanied by examples from previous projects. These remarks indicate that students not only recognized the value of the Problem Vision board but also envisioned its application in broader academic or professional contexts.
    
\subsection{Advisors' Perceived the support provided by Lean Research Inception (RQ2)}
\begin{figure}[h!]
    \centering
    \includegraphics[width=0.8\linewidth]{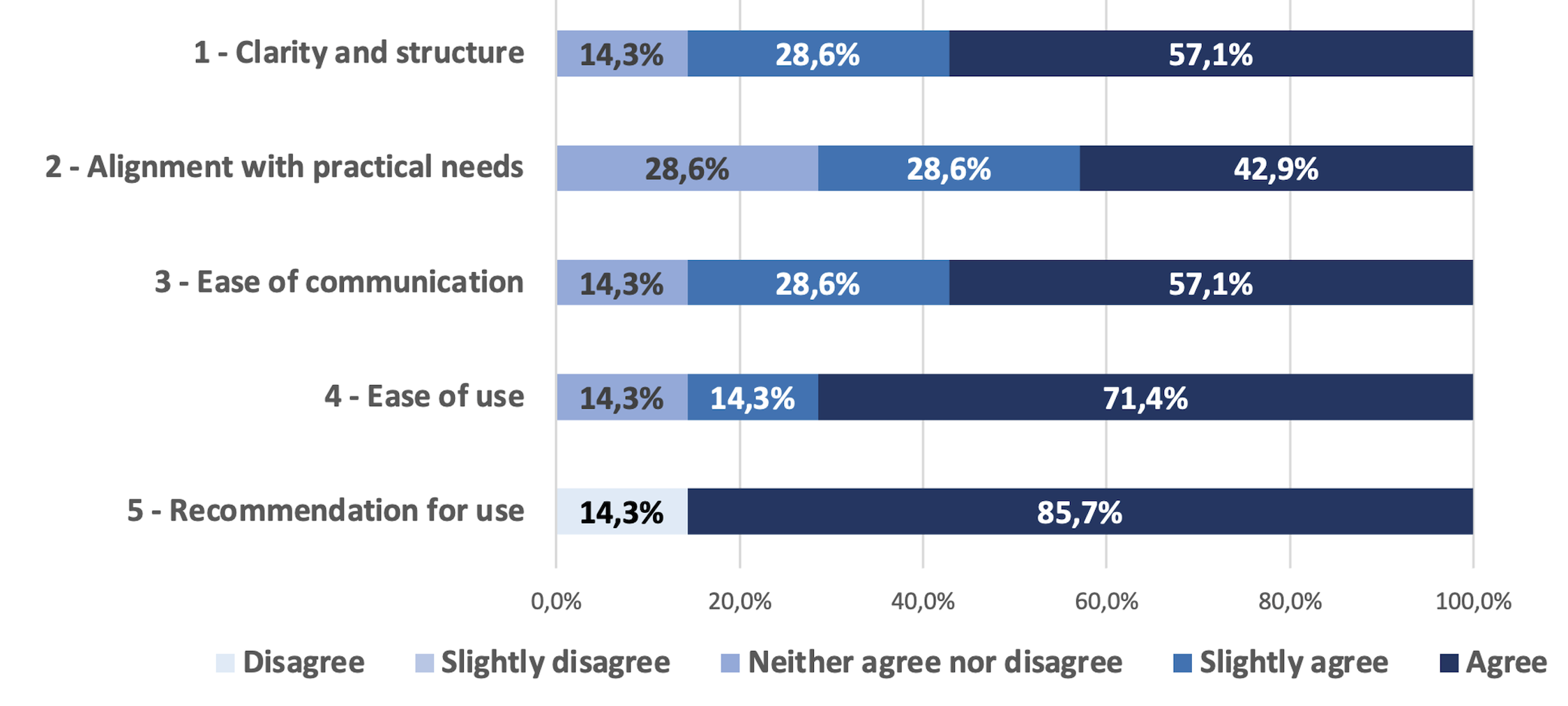}
    \caption{Results of the Advisors' Questionnaire (\%)}
    \label{fig:chart_advisors}
\end{figure}

The results of RQ2 focus on faculty advisors’ perceptions. The analysis addresses clarity and structure, alignment with practical needs, ease of communication, ease of use, and advisors’ intention to recommend the approach. Figure ~\ref{fig:chart_advisors} shows the results.

\textbf{Clarity and structure of the formulated research problems:} 57.1\% of the advisors agreed and 28,6\% partially agreed that research problems formulated using the Problem Vision board were clearer and better structured than usual, while only 14.3\% of the advisor expressed a neutral position. Qualitative feedback further supports this finding. Several advisors emphasized that the structured attributes of the Problem Vision board helped students articulate their ideas more clearly and systematically. For instance, P2 commented: \textit{``Excellent approach to supporting students, greatly facilitates the organization of ideas''}. highlighting the board’s role in structuring students’ reasoning. Similarly, P4 noted that the approach helped make students’ problems more concrete and less vague, facilitating more objective discussions during supervision. Some advisors also provided reflections that suggest opportunities for refinement. P1 mentioned that, although the board improves structure, students may still require additional guidance to deepen their understanding of the problem domain, indicating that the artifact is most effective when combined with active supervision. These comments suggest that while the Problem Vision board can enhance clarity and structure, its impact is further strengthened when used as part of a guided pedagogical process.

\textbf{Alignment with practical needs:} 42.9\% of the advisors agreed and 28.6\% partially agreed that the research problems formulated were better aligned with professional practice, while 28.5\% advisors expressed neutral perceptions. Several advisors highlighted that the Problem Vision board helps make gaps between students’ assumptions and real-world needs more explicit. For instance, P1 noted that \textit{``Students need a better foundation for thinking about the problem.''} indicating that the board exposes limitations in students’ understanding of practical contexts and facilitates more targeted guidance during supervision. Similarly, P3 mentioned that the structured discussion encouraged by the board helps challenge overly abstract or idealized problem statements, promoting a more realistic framing of the research problem. Some advisors implicitly suggested opportunities for strengthening this dimension. Comments indicate that the board is particularly effective when advisors actively leverage it to question evidence, stakeholders, and context attributes. These observations suggest that, while the Problem Vision board supports alignment with practical needs, its full potential is realized when combined with domain-specific feedback and examples drawn from real-world scenarios.

\textbf{Ease of communication of the research problem:} 57.1\% of the advisors agreed and 28.6\% partially agreed that the board facilitated the communication of students’ project ideas, while 14.3\% expressed a neutral perception. Advisors highlighted that the board helped make students’ reasoning more explicit, reducing ambiguities during conversations. For example, P2 commented that the approach \textit{``It greatly facilitates the organization of ideas''} which, in turn, supports clearer communication between students and advisors. Similarly, P4 noted that having the problem visually structured made discussions more direct and productive, as key elements could be easily referenced and questioned during meetings. Some advisors implicitly pointed out that the board also supports communication by revealing weaknesses in students’ explanations. P1 mentioned that students often struggle with articulating the problem clearly, and that the board helps externalize these difficulties, making them easier to address during supervision. These observations indicate that the board functions not only as a presentation artifact, but also as a shared cognitive and communication tool.

\textbf{Ease of use for students:} 71.4\% advisors agreed, 14.3\% partially agreed, and 14.3\% expressed a neutral perception. P7 explicitly stated: \textit{``Easy-to-understand framework''}, highlighting the accessibility and intuitiveness of the board. Other advisors noted that students were able to engage with the board with minimal difficulty, which allowed supervision sessions to focus more on the substance of the problem rather than on explaining the tool itself. Some advisors also hinted at opportunities for minor refinements. For instance, comments suggest that students with less prior experience in research might benefit from additional examples or brief guidance during the initial use of the board. These observations indicate that, while the Problem Vision board is generally easy to use, its effectiveness can be further enhanced through complementary instructional support.

\textbf{Intention to recommend the approach:} 
85,7\% faculty advisors stated that they would recommend the use of the Problem Vision board in other SE projects. Several advisors did not identify relevant limitations or needed improvements. For instance, P3 stated: 
\textit{``None so far''}, when asked about possible improvements, suggesting overall satisfaction with the approach. Other comments throughout the questionnaire emphasize that the board effectively supports structuring, communication, and guidance, which likely contributes to advisors’ willingness to recommend its use.

\section{Discussion}
The results of this study can reinforce the role of Lean Research Inception (LRI) as a pedagogical approach capable of supporting the teaching of practice-aligned research problem formulation in the context of Capstone Project courses in Software Engineering (SE). In contrast to approaches that emphasize project execution or the teaching of empirical methods~\cite{vilela2024empowering, greco2025learning}, LRI directly addresses a gap identified in the literature: structured support for the early stage of research problem formulation~\cite{garousi2020}, particularly in educational contexts such as Capstone courses~\cite{molleri2024teaching}.

From the students’ perspective (RQ1), the data indicate that LRI can foster reflection, clarity, and structured reasoning around the research problem. These findings align with initiatives that employ supporting artifacts, such as guides and templates, to facilitate the organization of ideas~\cite{molleri2024teaching}. The Problem Vision board functioned as a cognitive artifact that helped students structure initially vague or fragmented proposals, promoting more critical and contextualized thinking. It is important to emphasize that for many, this was their first experience with the structured formulation of a research problem. This is especially relevant given that prior studies continue to report challenges related to critical thinking development and clarity of objectives in research methods courses~\cite{greco2025learning}. Furthermore, the results suggest that LRI can complement educational models based on active learning and industry partnerships~\cite{romao2024agile} by providing a clear initial structure for students to approach real-world problems in a more grounded manner. The intention to adopt the approach in future contexts, expressed by most students, reinforces the transferable potential of LRI beyond the specific course setting.

Regarding the communication dimension, the data revealed variability in students’ responses. While some recognized the value of the board as a facilitator of dialogue with their advisors, others pointed to the need for more explicit guidance on its collaborative use. This finding suggests opportunities for pedagogical refinement, particularly in courses that rely on remote supervision. Nevertheless, the use of the board as a mediating artifact during advisory meetings aligns with studies that emphasize the importance of supporting artifacts to enhance interaction between students and supervisors in project-based learning environments~\cite{da2024undergraduate}.

The perspective of faculty advisors (RQ2) further reinforces the benefits perceived by students. Advisors reported improvements in the clarity, structure, and practical relevance of the formulated problems, and recognized the role of LRI in making the supervision process more objective and productive, contributing to clearer discussions and more focused feedback, particularly in remote supervision contexts. This aspect aligns closely with university-industry collaboration initiatives aimed at bridging the gap between academic environments and real-world market demands~\cite{goelzer2025s}, as LRI helps make explicit students’ gaps in understanding the practical context, thereby fostering more critical discussions around stakeholders, evidence, and implications.

Comparing students’ and advisors’ perceptions reveals some differences. While both groups reported positive views of the framework, ease of use was perceived more strongly by advisors (71.4\%) than by students (60\%). A possible explanation is that advisors can compare the approach with previous cohorts, for example, whereas students typically lack a similar comparison baseline. Finally, the strong intention among advisors to recommend the approach can reinforce its institutional acceptance and suggests that LRI can be incorporated as a pedagogical strategy across different educational contexts. By addressing in a structured manner a fundamental and often overlooked stage of the research process, LRI positions itself as an original and complementary contribution to existing educational practices in SE. These findings advance the direction proposed by~\cite{molleri2024teaching} by investigating how pedagogical structures can support the articulation between methodological rigor and practical relevance in SE education.

From a pedagogical perspective, LRI can support SE students by helping them transform vague ideas into clearer, more structured, and practice-aligned research problems. By using the Problem Vision board, students are encouraged to explicitly articulate attributes such as \textit{context}, \textit{stakeholders}, \textit{evidence}, and \textit{objectives}, which helps organize their reasoning and promotes more systematic reflection. From an instructional perspective, LRI can be integrated into Capstone courses as an early-stage activity to support research problem definition. Using the framework during the first weeks of the course and throughout supervision meetings can foster more structured discussions between students and advisors, enabling the identification of weaknesses in problem formulation at early stages of the project. %In this way, the approach can help students develop a more structured and reflective way of formulating research problems in SE.

\section{Limitations and Threats to Validity}
We addressed the four categories of validity threats~\cite{wohlin2024}:

\textbf{Internal Validity:} Internal validity may be affected by the study’s educational context. As the use of the Problem Vision board was part of a pedagogical initiative, participants’ responses may have been influenced by social desirability bias or by the hierarchical relationship between students and faculty advisors. Additionally, differences in advisors’ experience, supervision style, and availability may have influenced how the approach was applied during supervision meetings. To mitigate these effects, participation was voluntary and responses had no impact on students’ academic evaluation.

\textbf{External Validity:} External validity is limited to a single undergraduate SE program at a Brazilian university, despite involving three campuses and a distance learning format. Although the same materials and guidelines were used, differences in local dynamics, advisor availability, or student profiles may exist. As no campus-level comparison was performed, potential contextual effects could not be analyzed, and the results may not generalize to other institutions, instructional formats, or academic levels.

\textbf{Construct Validity:} Threats to construct validity are related to the data collection instruments. Participants’ perceptions were captured through Likert-scale questionnaires, which may not fully represent the complexity of the experiences lived during the application of the approach. In addition, concepts such as clarity, contextualization, and communication may have been interpreted differently by participants. To mitigate these limitations, the questionnaires included open-ended questions, allowing qualitative elaboration and alignment with the study objectives and the core dimensions of LRI.

\textbf{Conclusion Validity:} The sample size and the descriptive nature of the analysis limit the use of inferential statistical techniques. The results are primarily based on descriptive statistics and qualitative analysis. To mitigate this limitation, quantitative and qualitative data triangulation was performed to support more robust interpretations. Future studies may adopt experimental or quasi-experimental designs, as well as larger samples, to strengthen the empirical evidence on the educational impact of LRI.

\section{Conclusion and Future Work}
This study investigated the use of Lean Research Inception (LRI) as a pedagogical strategy to support the formulation of practice-aligned research problems in SE education. Applied in Capstone Project courses, LRI, through its central artifact, the Problem Vision board, helped students reflect more effectively on their research problems (60\%), clarify and structure them more precisely (61.7\%), and communicate them more clearly (50\%). The results also indicate that LRI was perceived as easy to use (60\%) and valuable for future projects (58.3\%). Faculty advisors also reported greater clarity in the structure of the formulated problems (57.1\%), highlighting support for alignment with practical needs (42.9\%) and for communicating the problem during the supervision process. Ease of use was identified as one of the main strengths (71.4\%), and the high intention to recommend the approach (85.7\%) reinforces the educational potential of LRI.

By focusing on the often-overlooked phase of research problem formulation, this study complements existing approaches centered on project execution or empirical methods. By shaping how future researchers learn to formulate problems, LRI has the potential to influence educational outcomes and the long-term relevance of SE research. Future work could explore the application of all five LRI phases in educational settings, enabling a more complete integration of research planning, evaluation, and decision-making. Replication in different institutions and learning contexts can further validate and extend the applicability of the approach.

\section*{Artifacts Availability}
All materials are available in our open science repository \cite{zenodoRepository}.

\section*{Acknowledgements}
We thank CNPq (Grants 312275/2023-4 and 420191/2025-9), CAPES (Doctoral and Postdoctoral Institutional Programs), FAPERJ (Grant E-26/204.256/2024), Kunumi Institute, and FUSVE for their generous support. We also thank the students and professors of the University of Vassouras for their contributions to this study.

\bibliographystyle{sbc}
\bibliography{sbc-template}

@article{codabux2024software,
  title={The Software Industry-Academia Collaboration: Novelty and Practice},
  author={Codabux, Zadia and Fard, Fatemeh Hendijani},
  journal={IEEE Software},
  volume={42},
  number={3},
  pages={133--140},
  year={2024},
  publisher={IEEE}
}

@inproceedings{kalinowski2020lean,
  title={Lean r\&d: An agile research and development approach for digital transformation},
  author={Kalinowski, Marcos and Lopes, H{\'e}lio and Teixeira, Alex Furtado and da Silva Cardoso, Gabriel and Kuramoto, Andr{\'e} and Itagyba, Bruno and Batista, Solon Tarso and Pereira, Juliana Alves and Silva, Thuener and Warrak, Jorge Alam and others},
  booktitle={International Conference on Product-Focused Software Process Improvement},
  pages={106--124},
  year={2020},
  organization={Springer}
}

@inproceedings{da2024undergraduate,
  title={An undergraduate Software Engineering practice course: bridging the academia-industry gap},
  author={Paiva, Sofia Larissa Costa and de Souza, Adriana Silveira and de Oliveira, Juliano Lopes and Ramada, Mariana Soller and da Luz, Murilo Lopes},
  booktitle={Simp{\'o}sio Brasileiro de Engenharia de Software (SBES)},
  pages={410--421},
  year={2024},
  organization={SBC}
}

@inproceedings{vilela2024empowering,
  title={Empowering Undergraduates in Empirical Research Methods: an Experience Report},
  author={Vilela, J{\'e}ssyka and Silva, Carla},
  booktitle={Simp{\'o}sio Brasileiro de Engenharia de Software (SBES)},
  pages={543--553},
  year={2024},
  organization={SBC}
}

@inproceedings{goelzer2025s,
  title={Do’s and Don’ts of Partnering with Industry to Educate Software Engineering Students: Recommendations Based on a Teaching Experience},
  author={Goelzer, Natalya Marjana and Possamai, Pedro Portella and Marczak, Sabrina},
  booktitle={Simp{\'o}sio Brasileiro de Engenharia de Software (SBES)},
  pages={577--587},
  year={2025},
  organization={SBC}
}

@inproceedings{romao2024agile,
  title={Agile Minds, Innovative Solutions, and Industry-Academia Collaboration: Lean {R}\&{D} Meets Problem-Based Learning in Software Engineering Education},
  author={Romao, Lucas and Kalinowski, Marcos and Barbosa, Clarissa and Ara{\'u}jo, Allysson Allex and Barbosa, Simone DJ and Lopes, Helio},
  booktitle={Simp{\'o}sio Brasileiro de Engenharia de Software (SBES)},
  pages={346--356},
  year={2024},
  organization={SBC}
}

@inproceedings{greco2025learning,
  title={Learning through Practice: Teaching Empirical Software Engineering for Undergraduate Students},
  author={Greco, Ranya Duran and Desid{\'e}rio, Sofia Bento and Bandeira, Caio C{\'e}sar Sousa and dos Santos Marques, Anna Beatriz},
  booktitle={Simp{\'o}sio Brasileiro de Engenharia de Software (SBES)},
  year={2025},
  organization={SBC}
}

@incollection{molleri2024teaching,
  title={Teaching Research Design in Software Engineering},
  author={Moll{\'e}ri, Jefferson Seide and Petersen, Kai},
  booktitle={Handbook on Teaching Empirical Software Engineering},
  pages={71--100},
  year={2024},
  publisher={Springer}
}

@article{garousi2020,
  title={Practical relevance of software engineering research: synthesizing the community’s voice},
  author={Garousi, Vahid and Borg, Markus and Oivo, Markku},
  journal={Empirical Software Engineering},
  volume={25},
  pages={1687--1754},
  year={2020},
  publisher={Springer}
}

@inproceedings{petersen2024,
  title={Revisiting the construct and assessment of industrial relevance in software engineering research},
  author={Petersen, Kai and B{\"o}rstler, J{\"u}rgen and Ali, Nauman Bin and Engstr{\"o}m, Emelie},
  booktitle={Proceedings of the 1st IEEE/ACM International Workshop on Methodological Issues with Empirical Studies in Software Engineering},
  pages={17--20},
  year={2024}
}

@article{molleri2023,
  title={Determining a core view of research quality in empirical software engineering},
  author={Moll{\'e}ri, Jefferson Seide and Mendes, Emilia and Petersen, Kai and Felderer, Michael},
  journal={Computer Standards \& Interfaces},
  volume={84},
  pages={103688},
  year={2023},
  publisher={Elsevier}
}

@book{plattner2009,
  title={Design thinking},
  author={Plattner, Hasso and Meinel, Christoph and Weinberg, Ulrich},
  year={2009},
  publisher={Springer}
}

@article{caroli2017,
  title={Lean inception},
  author={Caroli, Paulo},
  journal={S{\~a}o Paulo, BR: Caroli. org},
  year={2017}
}

@book{wohlin2024,
  author       = {Claes Wohlin and
                  Per Runeson and
                  Martin H{\"{o}}st and
                  Magnus C. Ohlsson and
                  Bj{\"{o}}rn Regnell and
                  Anders Wessl{\'{e}}n},
  title        = {Experimentation in Software Engineering, Second Edition},
  publisher    = {Springer},
  year         = {2024}
}

@article{basili1988,
  title={The TAME project: Towards improvement-oriented software environments},
  author={Basili, Victor R and Rombach, H Dieter},
  journal={IEEE Transactions on software engineering},
  volume={14},
  number={6},
  pages={758--773},
  year={1988},
  publisher={IEEE}
}

@book{runeson2012case,
  title={Case study research in software engineering: Guidelines and examples},
  author={Runeson, Per and Host, Martin and Rainer, Austen and Regnell, Bjorn},
  year={2012},
  publisher={John Wiley \& Sons}
}

@article{zenodoRepository,
  title={Artifacts: Teaching Practically Relevant Research Problem Formulation in Software Engineering with Lean Research Inception},
  author={Pereira, Anrafel Fernandes and Ornelas, Tataine and  
Araújo, Allysson Allex and Kalinowski, Marcos},
  journal={https://doi.org/10.5281/zenodo.19074876},
  volume={},
  number={},
  pages={},
  year={2026},
  publisher={Zenodo}
}

@book{ries2011,
  title={The lean startup: How today's entrepreneurs use continuous innovation to create radically successful businesses},
  author={Ries, Eric},
  year={2011},
  publisher={Crown Currency}
}

@inproceedings{PaperEASE2025,
  author       = {Anrafel Fernandes Pereira and Marcos Kalinowski and Maria Teresa Baldassarre and 
                  J{\"u}rgen B{\"o}rstler and Nauman bin Ali and Daniel Mendez},
  title        = {Towards Lean Research Inception: Assessing Practical Relevance of Formulated Research Problems},
  booktitle    = {Evaluation and Assessment in Software Engineering (EASE '25)},
  year         = {2025},
  address      = {Istanbul, Turkiye},
  publisher    = {ACM}
}

@inproceedings{PaperWSESE2026,
  author       = {Anrafel Fernandes Pereira and Maria Teresa Baldassarre and Daniel Mendez and J{\"u}rgen B{\"o}rstler and Nauman bin Ali and Rahul Mohanani and Darja Smite and Stefan Biffl and Rogardt Heldal and Davide Falessi and Daniel Graziotin and Marcos Kalinowski},
  title        = {Attributes to Support the Formulation of Practically Relevant Research Problems in Software Engineering},
booktitle    = {3rd International Workshop on Methodological Issues with Empirical Studies in Software Engineering (WSESE@ICESE'26)},
  year         = {2026}
}

\end{document}